# 3D MR Fingerprinting for Dynamic Contrast-Enhanced Imaging of Whole Mouse Brain

Yuran Zhu[1], Guanhua Wang[2], Yuning Gu[1], Walter Zhao[1], Jiahao Lu[1], Junqing Zhu[3], Christina J. MacAskill[1], Andrew Dupuis[1], Mark A. Griswold[1,3], Dan Ma[1,3], Chris A. Flask[1,3,4], Xin Yu[1,3,5]

[1]Department of Biomedical Engineering, Case Western Reserve University, Cleveland, Ohio, USA

[2]Department of Biomedical Engineering, University of Michigan, Ann Arbor, Michigan, USA

[3]Department of Radiology, Case Western Reserve University, Cleveland, Ohio, USA

[4]Department of Pediatrics, Case Western Reserve University, Cleveland, Ohio, USA

[5]Department of Physiology and Biophysics, Case Western Reserve University, Cleveland, Ohio, USA

Address correspondence to:

Xin Yu, Sc.D.,
Wickenden 430
10900 Euclid Avenue
Cleveland, OH 44106, USA
Tel: 216-368-3918
Email: xin.yu@case.edu

# Abstract

**Purpose:** Quantitative MRI enables direct quantification of contrast agent concentrations in contrast-enhanced scans. However, the lengthy scan times required by conventional methods are inadequate for tracking contrast agent transport dynamically in mouse brain. We developed a 3D MR fingerprinting (MRF) method for simultaneous $T_1$ and $T_2$ mapping across the whole mouse brain with 4.3-min temporal resolution.

**Method:** We designed a 3D MRF sequence with variable acquisition segment lengths and magnetization preparations on a 9.4T preclinical MRI scanner. Model-based reconstruction approaches were employed to improve the accuracy and speed of MRF acquisition. The method's accuracy for $T_1$ and $T_2$ measurements was validated in vitro, while its repeatability of $T_1$ and $T_2$ measurements was evaluated in vivo (n=3). The utility of the 3D MRF sequence for dynamic tracking of intracisternally infused Gd-DTPA in the whole mouse brain was demonstrated (n=5).

**Results:** Phantom studies confirmed accurate $T_1$ and $T_2$ measurements by 3D MRF with an undersampling factor up to 48. Dynamic contrast-enhanced (DCE) MRF scans achieved a spatial resolution of $192 \times 192 \times 500$ µm$^3$ and a temporal resolution of 4.3 min, allowing for the analysis and comparison of dynamic changes in concentration and transport kinetics of intracisternally infused Gd-DTPA across brain regions. The sequence also enabled highly repeatable, high-resolution $T_1$ and $T_2$ mapping of the whole mouse brain ($192 \times 192 \times 250$ µm$^3$) in 30 min.

**Conclusion:** We present the first dynamic and multi-parametric approach for quantitatively tracking contrast agent transport in the mouse brain using 3D MRF.

## 1. Introduction

Magnetic resonance imaging (MRI) of mouse brain provides invaluable insights into the pathophysiology of neurological disorders and contributes to the advancement of novel therapeutic interventions.[1] Among the tissue parameters measurable by MRI, the longitudinal ($T_1$) and transverse ($T_2$) relaxation times are of particular interest due to their sensitivity to pathological changes within the tissue, allowing them to serve as critical biomarkers.[2] Measuring $T_1$ and $T_2$ by quantitative MR (qMR) methods provides standardized and reproducible evaluations of disease progression and response to therapy.[3] Furthermore, in cases of dynamic contrast-enhanced (DCE) imaging, quantifying $T_1$ and $T_2$ changes enables tracking of concentration changes of the contrast agents, thereby supporting pharmacokinetic modeling of tracer dynamics within the system.[4,5] However, the conventional approaches cannot dynamically measure contrast agent transport across the whole mouse brain with sufficient spatial and temporal resolution.[6–8]

Magnetic resonance fingerprinting (MRF) has emerged as an accurate and efficient method for simultaneously mapping multiple tissue parameters in a single scan with unprecedented speed.[9] Although now widely implemented and evaluated in human studies,[10] few studies have been conducted to exploit MRF's potential as an investigative tool for basic science studies in mouse.[11–14] The development of MRF for imaging mouse brain has been greatly challenged by the significantly lower signal-to-noise ratio (SNR) due to the submillimeter voxel size needed to delineate mouse brain.[15] Existing two-dimensional (2D) preclinical MRF sequences have proven to be fast and effective for studying healthy and diseased mouse brains and other tissues.[11,12] These 2D MRF methods have also facilitated the dynamic quantification of contrast agent transport in tumor-bearing mouse models.[12,16] However, the three-dimensional (3D) implementations have suffered from unacceptably prolonged scan time. A previous attempt to map whole-brain $T_1$ and $T_2$ in rat required 1.5 hrs to achieve a 0.5-mm isotropic resolution with a Cartesian trajectory, rendering the assessment of contrast agent dynamics unfeasible.[14] Similarly, we recently implemented a stack-of-spirals trajectory to achieve a spatial resolution of 0.35 × 0.35 × 1 mm$^3$ for simultaneous $T_1$ and $T_2$ mapping across the entire macaque brain at 9.4 T in 1.8 hrs.[17] While the study showed promise, further improvement in temporal and spatial resolutions is necessary to allow for efficient multiparametric mapping of the entire mouse brain.

In this study, we developed a 3D MRF sequence for simultaneous multiparametric mapping of the whole mouse brain with unparalleled speed and spatial resolution. The sequence acquired different numbers of time frames following the inversion and $T_2$ preparations to more efficiently capture image contrast generated by the preparation modules. Iterative image reconstruction by subspace-based low-rank modeling was implemented on graphic processing units (GPUs) to improve the accuracy and efficiency of multiparametric mapping. The MRF sequence's accuracy in estimating $T_1$ and $T_2$ was validated through in vitro phantom studies. This method enabled $T_1$ and $T_2$ mapping in whole mouse brain at a spatial resolution of 192 × 192 × 250 µm$^3$ in 30 min. We also demonstrated its application in dynamically and quantitatively tracking the transport of a gadolinium-based contrast agent (GBCA) in cerebrospinal fluid (CSF) after its intracisternal infusion, where 32 MRF scans of 4.3-min temporal resolution and 192 × 192 × 500 µm$^3$ spatial resolution were collected.

## 2. Methods

### 2.1 MRF pulse sequence and data acquisition

The MRF sequence was partitioned into 8 segments with different magnetization preparations to encode $T_1$, $T_2$, and $M_0$ sensitivity (Figure 1). Specifically, an adiabatic inversion pulse was applied before segments 1 and 5, with an inversion time of 21 ms to enhance $T_1$ sensitivity. $T_2$-preparation modules were applied before segments 3, 4, 7, and 8, with varying mixing times of 25 ms, 50 ms, 75 ms, and 100 ms, respectively.[17,18] The acquisition of varying numbers of time frames in each segment was tailored to more effectively capture the $T_1$, $T_2$, and $M_0$ contrasts produced by the preparation module preceding the segment. A total of 768 frames were acquired using fast imaging with steady-state precession (FISP) scheme with a constant repetition time (TR) of 10 ms, a constant echo time (TE) of 2 ms, and a dephasing gradient of 4π.[19] All excitations used a 1.5-ms sinc pulse with 7 lobes. A varying flip angle (FA) pattern with low nominal angles (<15°) was used to reduce the influences of $B_1$ inhomogeneity.[20] A 300-ms delay was applied between two consecutive acquisition segments and a 3-s delay was applied after one fingerprint acquisition to allow partial magnetization recovery. The acquisition time of a single fingerprint was ~13 s.

A stack-of-spirals trajectory was used to acquire the 3D MRF data.[21] Two uniform density spiral (UDS) trajectories with a readout bandwidth of 200 kHz were designed for in vitro and in vivo studies, respectively. For in vitro studies on phantoms, a field of view (FOV) of 50 × 50 mm$^2$ was sampled with a matrix size of 256 × 256, leading to an in-plane resolution of 195 × 195 µm$^2$. For in vivo studies, FOV and matrix size were 30 × 30 mm$^2$ and 156 × 156, respectively, yielding an in-plane resolution of 192 × 192 µm$^2$. The readout time of a single spiral interleaf was 4.1 and 2.9 ms for in vitro and in vivo scans, respectively. All interleaves of both spiral trajectories were measured on a water phantom.[22]

The spiral trajectories fully sampled the k-space with 48 spiral arms, and the rotation angle between two adjacent spiral arms was 7.5° (360°/48). Acceleration of data acquisition was achieved by undersampling in both the in-plane ($k_{xy}$) and through-plane ($k_z$) directions as described previously.[17,21] For scans with an in-plane undersampling factor of $R_{xy}$, 48/$R_{xy}$ spiral arms with a rotation angle of $R_{xy}×7.5°$ were sampled in each frame. Within each fingerprint acquisition, the spiral arms were rotated by 7.5° from one frame to the next. For the acquisition of $N_s$ slices with a through-plane undersampling factor of $R_z$, the phase-encoding ($k_z$) lines were divided into $N_s/R_z$ partitions with $R_z$ lines in each partition. Undersampling was achieved by acquiring one fingerprint from each partition and repeatedly cycling the acquisition of the $R_z$ lines within the same partition across all the time frames.

## 2.2 Dictionary simulation

A dictionary of simulated fingerprints was generated by solving the Bloch equation using the acquisition parameters of the MRF sequence. Specifically, nominal flip angles were used in simulating the excitation. To account for the 4π dephasing at the end of each TR, a total of 81 isochromats with dephasing evenly distributed from -2π to 2π were simulated and averaged. Variable step sizes were used to generate the $T_1$ and $T_2$ entries for the dictionary, where the $T_1$ and $T_2$ values were more densely distributed in the range of brain parenchyma in pre- and post-contrast conditions.[12,17] In particular, a 20-ms step size was used for $T_1$ and a 0.5-ms step size was used for $T_2$ at the more dense regions. $T_1$ step size incremented to 200 ms beyond 2000 ms. The $T_2$ step size was adjusted to 2 ms for values between 40 and 60 ms, increased to 10 ms for values between 60 and 200 ms, and set to 20 ms for values over 200 ms. A total of 126 $T_1$ values from 100 ms to 8000 ms and 111 $T_2$ values from 2 ms to 400 ms were used, leading to 13,986 dictionary entries in total.

## 2.3 Image reconstruction

Iterative image reconstruction was employed with low-rank (LR) subspace modeling to mitigate aliasing artifacts from undersampling.[23] The MRF data, as a time series of images, can be represented as a Casorati matrix ($C^{N×M}$), with spatial and temporal dimensions represented by its rows ($N$) and columns ($M$), respectively. A low-rank constraint can be enforced on $C$ to exploit the strong spatial-temporal correlation within the acquired MRF data. Specifically, a Casorati matrix of rank $L$ ($L \ll N, M$) can be factorized into two matrices $U^{N \times L}$ and $V^{L \times M}$, i.e., $C = UV$, with the rows of $U$ and columns of $V$ span the spatial and temporal subspaces of $C$, respectively.[24] The image reconstruction for MRF can thus be estimated iteratively by minimizing the following cost function

$$\widehat{U} = arg\ min_U \left\|AU\widehat{V} - y\right\|_2^2 + \lambda\phi(U)$$

where $A$ represents the system encoding matrix that spans all receiver channels, MRF time frames, and sampled k-space dimensions within each time frame. $\phi(U)$ denotes an L1 regularizer, i.e., $\phi(U) = \left\|U\right\|_1$, with the weight $\lambda$ chosen from a grid search. The temporal pre-estimation, $\widehat{V}$, was extracted from the singular value decomposition (SVD) of the simulated MRF dictionary. The rank of the model ($L$) was heuristically determined, and a rank of 12 reflected the low-rank constraint that the singular value of the dictionary had decayed by >90%.

All image reconstruction and dictionary matching were performed using PyTorch on an Nvidia Tesla V100 GPU. Image reconstruction incorporated measured spiral trajectories and coil sensitivity maps calculated from the time-series-compressed MRF data by ESPIRiT.[25,26] The conjugate gradient (CG) and Fast Iterative Shrinkage-Thresholding Algorithm (FISTA) were employed to iteratively solve the smooth ($\lambda = 0$) and non-smooth ($\lambda > 0$) cost functions, respectively.[27,28] The convergence was empirically determined based on the residual norm ($\left\|AU\widehat{V} - y\right\|/\left\|y\right\|$).[29] The LR-based image reconstruction method was compared with the conventional reconstruction through the inverse non-uniform fast Fourier transform (iNUFFT).[30] Both conventional and LR-based image reconstructions were executed using an open-source toolkit, MIRTorch.[31,32] Leveraging Toeplitz embedding and data parallelism, the reconstruction time of the LR algorithm was significantly accelerated for each 3D MRF scan on our current hardware.[33,34]

**2.4 In vitro validation in phantom**

All MRI studies were performed on a 9.4T Bruker BioSpin system operating on ParaVision 360 v3.3 platform with a maximal gradient of 654 mT/m (Bruker Biospin, Billerica, MA). A 6-compartment phantom containing different concentrations of manganese chloride ($MnCl_2$, 50 to 300 µM) was used to mimic the $T_1$ and $T_2$ values in brain parenchyma at 9.4 T.[12,17] A 40-mm 2-channel transmit/receive volume coil was used for data acquisition. MRF data were acquired from 20 slices with an in-plane resolution of 195 $\times$ 195 µm$^2$ and a thickness of 0.5 mm. The acquisition of a fully sampled dataset was about 3 hr 28 min. Retrospective undersampling was performed on the fully sampled MRF data to evaluate different undersampling schemes, and the accuracies of $T_1$ and $T_2$ estimation were compared against the results of fully sampled data. $T_1$ and $T_2$ maps of the center slice were also acquired using conventional methods, with a lower in-plane resolution of 468 × 468 μm$^2$ and a slice thickness of 1 mm. Specifically, $T_1$ was measured using the inversion-recovery spin-echo (IR-SE) method with 15 inversion times (TIs) ranging from 10 ms to 6 s. TR and TE were 10 s and 4.6 ms, respectively. $T_2$ was measured using the spin-echo method with 15 TEs ranging from 4 ms to 800 ms and a constant TR of 10 s.

**2.5 In vivo experiments**

The animal protocol was approved by the Institutional Animal Care and Use Committee (IACUC) of Case Western Reserve University. All experiments were performed on male C57BL/6 mice (Jackson Laboratories, Bar Harbor, ME, USA). The animals were housed in a temperature- and humidity-controlled environment with ad libitum access to food and water and a 12-h light-dark cycle. Anesthesia was maintained with 1-1.5% isoflurane in 100% $O_2$ during all surgical procedures and MRI scans. All MRI acquisitions used an 86-mm volume coil for transmitting and a 4-channel array coil for receiving. Respiration rate and body temperature were monitored during all MRI experiments. The respiratory rate was maintained within 90 to 110 breaths per minute (bpm) by adjusting the anesthesia level. The body temperature was maintained at ~37°C by blowing warm air into the scanner through a feedback control system (SA Instruments, Stony Brook, NY, US).

$T_1$ and $T_2$ maps of the whole mouse brain were acquired by 3D MRF with 24-fold in-plane undersampling (n=3). Imaging parameters were: FOV, 30 × 30 × 17.5 mm$^3$; matrix size, 156 × 156 × 70; slice thickness, 0.25 mm; in-plane resolution, 192 × 192 μm$^2$. Total scan time was ~30 min. $T_2$-weighted ($T_2$w) images with matching FOV and spatial resolution were also acquired using a multislice RARE sequence with a TR of 4000 ms and an effective TE of 25.8 ms. The MRF measurements of $T_1$ and $T_2$ were repeated on different days to evaluate the reproducibility of the proposed method. All mice were 14 weeks old, and the average body weight was 31.1 g at the time of MRI experiments.

Dynamic MRF scans were performed on mice with intracisternal infusion of Gd-DTPA (n = 5). Intracisternal cannulation of a polyethylene micro-tubing (0.13 mm ID × 0.25 mm OD, Scientific Commodities, Lake Havasu City, AZ, USA) was performed before MRI scans as previously described.[35] All mice were 10-12 weeks old at the time of the experiment and had an average body weight of 27.1 g. An undersampling factor of 48 ($R_{xy}$=24, $R_z$=2) was used for dynamic mapping with a temporal resolution of 4 min 20 s. A total of 20 sagittal slices of 0.5-mm thickness that covered the whole brain were acquired, with the same in-plane resolution as used in the anatomical mapping (192 × 192 µm$^2$). One MRF dataset was acquired at baseline before Gd-DTPA infusion, followed by 31 MRF scans, for a total scan time of 135 min. A total of 10 µL Gd-DTPA of 12.5 mM concentration was infused at a rate of 0.33 µL/min for 30 min after the baseline MRF acquisition.

### 2.6 Data analysis

All image and data analyses were performed using either in-house developed or open-source software in MATLAB (MathWorks, Natick, MA, USA) or Python (Python Software Foundation, v.3.0). All MRF maps were reconstructed using the LR method. For the repeatability study, MRF-derived parameter maps from two different days were skull stripped and co-registered to an MRI mouse brain atlas through affine and deformable transformations using the Advanced Normalization Tools (ANTs).[36–39] Voxel-wise correlation analyses that included the entire brain were performed to evaluate the reproducibility of in vivo $T_1$ and $T_2$ measurements. Furthermore, mean $T_1$ and $T_2$ values in 13 regions of interest (ROIs) were calculated. For dynamic study, motion correction was performed by registering the dynamic maps to the baseline via affine transformation. 11 ROIs at major anatomical landmarks were manually selected. Dynamic changes in relaxation rates from baseline ($\Delta R_1$ and $\Delta R_2$) were calculated from the dynamic $T_1$ and $T_2$ maps. Subsequently, mean dynamic curves were obtained without normalization.

## 3. Results

### 3.1 $T_1$ and $T_2$ validation in phantom

The accuracy of $T_1$ and $T_2$ measurements from MRF scans at different undersampling factors was first validated in the multi-compartment phantom with iNUFFT-based image reconstruction (Figure S1). The compartment-wise mean $T_1$ and $T_2$ values measured at R = 48 from the center slice were in excellent agreement with the IR-SE and SE measurements from the same slice location, respectively (Figure 2c,f). Furthermore, inter-slice variations in MRF-measured $T_1$ and $T_2$ values were insignificant (Figure S2), suggesting that the FISP-based MRF method is insensitive to $B_0$ inhomogeneity. The $R_1$ and $R_2$ values derived from 48-fold undersampled MRF also demonstrated a linear relationship to $Mn^{2+}$ concentrations and the calculated relaxivities were in strong agreement with those derived from IR-SE or SE methods, respectively (Figure S3).

LR reconstruction mitigated the aliasing artifacts from undersampling (Figure S4), leading to reduced compartment-wise standard deviations of $T_1$ and $T_2$ estimations at R = 48 (Figure 2a,d). The normalized root-mean-squared error (NRMSE) of MRF-derived $T_1$ and $T_2$ maps was evaluated at different undersampling factors for both iNUFFT- and LR-based image reconstructions (Figure 2b,e). Both $T_1$ and $T_2$ NMRSEs were under 10% at R = 48 when reconstructed with LR as compared to fully sampled MRF. The rate of residual decay was evaluated on the 48-fold undersampled phantom data across 50 iterations of CG (Figure S5). Significant convergence was achieved by 20 iterations, with the subsequent iterations showing minimal residual decay. The entire LR-based reconstruction pipeline runs in under 20 min, thus enabling the efficient reconstruction of dynamic MRF scans. Based on these results, 48-fold undersampling using LR reconstruction was determined to be both accurate and efficient for DCE-MRF, leading to a temporal resolution of 4 min 20 s for dynamic, simultaneous $T_1$ and $T_2$ mapping at a spatial resolution of $192 \times 192 \times 500$ μm$^3$.

**3.2 Mapping of $T_1$ and $T_2$ in the whole brain**

$T_1$ and $T_2$ maps acquired with a slice thickness of 0.25 mm and 24-fold undersampling are shown in Figures 3-4, along with $T_2$w images for comparison of anatomic structures. The image reconstruction time was approximately 30 min. The excellent delineation of major anatomic structures can be appreciated in all three views. In particular, MRF was able to differentiate gray and white matter across the whole brain, and delineate small structures such as the horns of the anterior commissure and corpus callosum (Figure 3). Fine anatomical details such as the corpus callosum are also discernible from the coronal and sagittal views (Figure 4). Major cerebral arteries were also notable on these $T_1$ and $T_2$ maps (Figure 4b), although there was an underestimation in measured $T_1$ and $T_2$ values of the blood due to flow effects.[40]

Figure 5 shows MRF-measured $T_1$ and $T_2$ values from 13 regions. Among all parenchymal regions analyzed, the internal capsule exhibited the lowest average $T_1$ (1453 ± 109 ms) and $T_2$ (22.1 ± 1.9 ms) values, while olfactory bulbs showed the highest ($T_1$ = 1914 ± 150 ms; $T_2$ = 33.2 ± 3.3 ms). Furthermore, the $T_1$ and $T_2$ values of CSF in the ventricular spaces were 3091 ± 168 ms and 143.8 ± 17.2 ms, respectively.

The repeatability of MRF measurements was evaluated in vivo from two scans performed on two different days (n = 3). Matching anatomical details can be appreciated in all three orthogonal views from co-registered maps (Figure 6a-b). The voxel-wise correlation coefficients ($r$) between the $T_1$ and $T_2$ maps from the two scans exceeded 0.96, suggesting high repeatability (Figure 6c-d).

### 3.3 Quantitative and dynamic mapping of Gd-DTPA transport in CSF

Figures 7 and 8 show dynamic changes in $T_1$ and $T_2$ before, during, and after Gd-DTPA infusion. The 3D MRF scans, with a temporal resolution of 4 min 20 s, effectively captured the dynamic changes induced by the intracisternal Gd-DTPA infusion in the whole brain, allowing quantitative tracking of contrast agent transport over time. The total reconstruction time for an entire dynamic dataset of 32 time points was approximately 8 hrs.

Gd-DTPA infusion induced a substantial decrease in $T_1$ (>1 s) in ROIs proximal to the infusion site, including cisterna magna, cerebellum, and the lower brain stem (Figure 7b). The $T_2$ changes in these regions were relatively small. While the cerebellum and lower brain stem experienced fast Gd-DTPA transport kinetics and high peak Gd-DTPA concentrations, the ΔR$_1$ increase in colliculi was significantly delayed, indicating an impediment in the transport of Gd-DTPA to this midbrain region from cisterna magna. In the meantime, Gd-DTPA transport along the ventral brain surface is prominent in both the $T_1$ maps and the time courses of ΔR$_1$ changes in these regions (Figure 7c). ROIs along the ventral brain surface showed a progressive increase in Gd-DTPA concentration accompanied by a reduction in peak concentration from posterior to anterior directions. Delayed transport kinetics and a 10-fold smaller Gd-DTPA magnitude can also be appreciated in the ROIs along the ventral brain surface as compared to those adjacent to the infusion site. The change in Gd-DTPA concentration reflected by both ΔR$_1$ and ΔR$_2$ changes were negligible in dorsal and deep brain regions except for hippocampus, suggesting that they are downstream of the Gd-DTPA transport (Figures 7d and 8d).

Notably, the time courses of ΔR$_1$ and ΔR$_2$ were consistent among individual animals across different brain regions (Figures S6 and S7). The mean time courses were calculated without any signal normalization, ensuring that the characteristic relaxation (ΔR$_1$ and ΔR$_2$) responses were identified for each region.

## 4. Discussion

We present the first 3D MRF method to dynamically track the transport of Gd-DTPA in the mouse brain through simultaneous multi-parametric mapping. The MRF sequence presented in the current work was designed to accurately map $T_1$ and $T_2$ in both pre- and post-contrast mouse brains during DCE studies. The acquisition efficiency of the current sequence was significantly improved compared to that of previous preclinical MRF studies. We also exploited a subspace-based, low-rank reconstruction approach to improve the accuracy of parameter mapping. The accuracy of $T_1$ and $T_2$ mapping, validated by undersampled phantom data, was in agreement with the gold-standard methods. The highly reproducible results enabled fast, accurate, and simultaneous in vivo $T_1$ and $T_2$ measurements in small laboratory animals. With a 48-fold undersampling, the MRF method allowed for dynamic tracking of the transport of intracisternally administered Gd-DTPA throughout the whole mouse brain, at a temporal resolution of 4 min 20 s and a spatial resolution of $192 \times 192 \times 500$ µm$^3$.

The 3D FISP-MRF method developed in the current study was designed with small flip angles (<15°) and did not include $B_0$ or $B_1$ variations in the dictionary simulation. Previous studies on human subjects have demonstrated that FISP-MRF sequences are immune to $B_0$ inhomogeneity.[19] Furthermore, FISP-MRF sequences with small flip angles are $B_1$-insensitive and thus do not require $B_1$ correction.[20] The considerably smaller FOV for mouse brain imaging and the use of an 86-mm volume coil for RF transmission also render $B_0$ and $B_1$ corrections unnecessary. However, in the case of a larger FOV and using a surface coil for transmission, $B_1$ correction may be necessary for improving the accuracy.[17]

To date, only a limited number of studies focused on mapping $T_1$ in human or animal brains at 9.4 T,[41–46] and even fewer studies measured $T_2$ at this field strength.[12,17,45] In cases where both $T_1$ and $T_2$ were of interest, sequential scans were performed to map $T_1$ and $T_2$ in the mouse brain using conventional methods, which led to a total acquisition time of over 1 hr for a single slice of the brain.[45] The current method enabled whole-brain $T_1$ and $T_2$ mapping in mouse brain with a spatial resolution of $192 \times 192 \times 250$ µm$^3$ in 30 min. $T_1$ and $T_2$ measured by 3D MRF were in good agreement with the previous studies using both MRF and non-MRF mapping methods. The MRF-measured $T_1$ ranged from 1300 to 1500 ms in white matter and 1800 to 2000 ms in gray matter, while the $T_2$ ranged from 20 to 30 ms in white matter and 30 to 40 ms in gray matter. Further, the $T_1$ (3091 ± 168 ms) and $T_2$ (143.8 ± 17.2 ms) of ventricular CSF quantified by the current MRF sequence were similar to those reported in a previous study of macaque brain at the same field strength using a different MRF sequence and image reconstruction method.[17,47] In comparison, a previous study employing a variable flip angle method reported $T_1$ values of about 2822 ms in CSF, 2001 ms in gray matter, and 1705 ms in white matter of rat brain.[41] Using the same method in mice, Xue and colleagues reported $T_1$ values of about 3000 ms in CSF and 2000 ms in normal brain tissue.[42] The MRF-measured $T_1$ values in the current study were generally in agreement with these previous studies using non-MRF methods.

The proposed MRF acquisition enabled dynamic and quantitative tracking of Gd-DTPA transport within the CSF pathway across the entire mouse brain. The observed time courses of $\Delta R_1$ and $\Delta R_2$ follow a linear relationship with the changes in Gd-DTPA concentration, enabling direct quantification of Gd-DTPA. This direct quantitative approach eliminates the need for signal normalization and reduces the impact of experimental variables such as animal positioning and coil loading. As a result, it leads to reduced signal variability when compared with previous studies employing simi-quantitative $T_1$-weighted methods.[33] Previous studies have employed conventional $T_1$ mapping methods to track intravenously or intrathecally administered GBCAs in mouse brain, where least-square curve fitting of image intensities was performed to derive dynamic $T_1$ values.[42,43,48] MRF offers a unique advantage that pattern matching between the acquired signal and the dictionary of simulated signals minimizes the effects of acquisition and undersampling errors while estimating multiple tissue parameters simultaneously.[9] Although we only demonstrated 3D DCE-MRF's application in tracking a single contrast agent in the current study, the multiparametric nature of the MRF method has the potential to be extended to simultaneously track two contrast agents.[16,49]

The 3D MRF method presented in the current study has significantly advanced whole-brain multiparametric mapping in mice. With a mouse brain about 2400 times smaller than the human brain,[50] preclinical MRF development has been challenged by the drastically lower SNR compared to human studies. This limitation was overcome by a synergistic approach of improved MRF sequence design and advanced reconstruction method employing low-rank subspace modeling. Iterative reconstruction on GPU power has substantially accelerated the image reconstruction process and facilitated the post-processing of DCE-MRF data, where a total of 32 time points (24,576 images) were acquired. Furthermore, simultaneous $T_1$ and $T_2$ mapping of the whole brain also provides the potential for simultaneous tracking of two MRI contrast agents with different relaxivities.[16,49] However, it is worth noting that while the 4.3-min temporal resolution of our DCE-MRF scans aligned well with the slow dynamics of CSF and was consistent with the current practices in the field, this temporal resolution is inadequate for dynamic tracking of a bolus injection of MRI contrast agent administered intravenously.

The spatial-temporal resolution of the current MRF method can be improved by further optimizing the sequence to shorten the acquisition time for each fingerprint,[51,52] as well as by optimizing the sampling trajectory to enhance efficiency.[53,54] Additionally, reconstruction methods that leverage signal redundancy along the dynamic dimension allow for further accelerated acquisition, and non-dictionary-based advanced reconstruction methods provide the potential for improve reconstruction.[47,55,56] These improvements hold promise for further improving the spatial and temporal resolution for dynamic mapping.

In conclusion, we present the first 3D MRF method for highly efficient and accurate $T_1$ and $T_2$ mapping of the entire mouse brain. This technique enables direct and quantitative tracking of contrast agent transport across various brain regions. Both acquisition and reconstruction efficiencies have been significantly improved compared to previous preclinical MRF studies. The proposed 3D MRF method holds the potential to greatly expand the capabilities of quantitative MRI in small animal imaging, with applications ranging from studies of CSF circulation to pharmacokinetics.

## Acknowledgments

This work made use of the High Performance Computing Resource in the Core Facility for Advanced Research Computing at Case Western Reserve University. This work was supported by grants from the National Institute of Health awards R01 NS124206 to X. Y., and predoctoral fellowship award from American Heart Association 23PRE1017924 to Y. Z. W.Z. was supported in part by NIH training grants T32 EB007509, T32 GM007250, and TL1 TR000441.

## Data availability statement

The experimental data, images, and code from this study are available upon request to the corresponding author.

**Figures and captions**

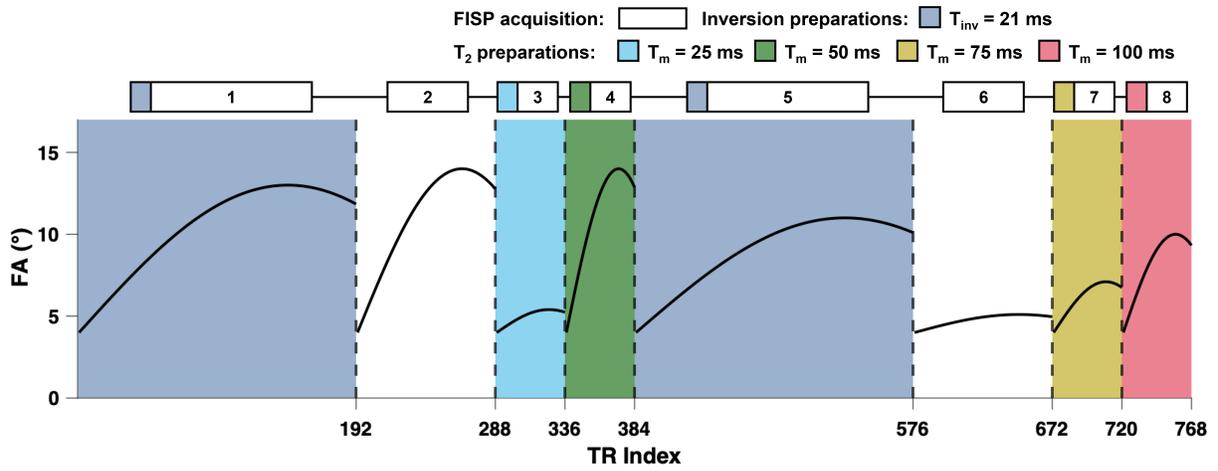

**Figure 1.** Schematics and flip angle pattern of the MR fingerprinting (MRF) sequence. Fast imaging with steady-state precession, FISP. Inversion delay, $T_{inv}$. Mixing time, $T_m$.

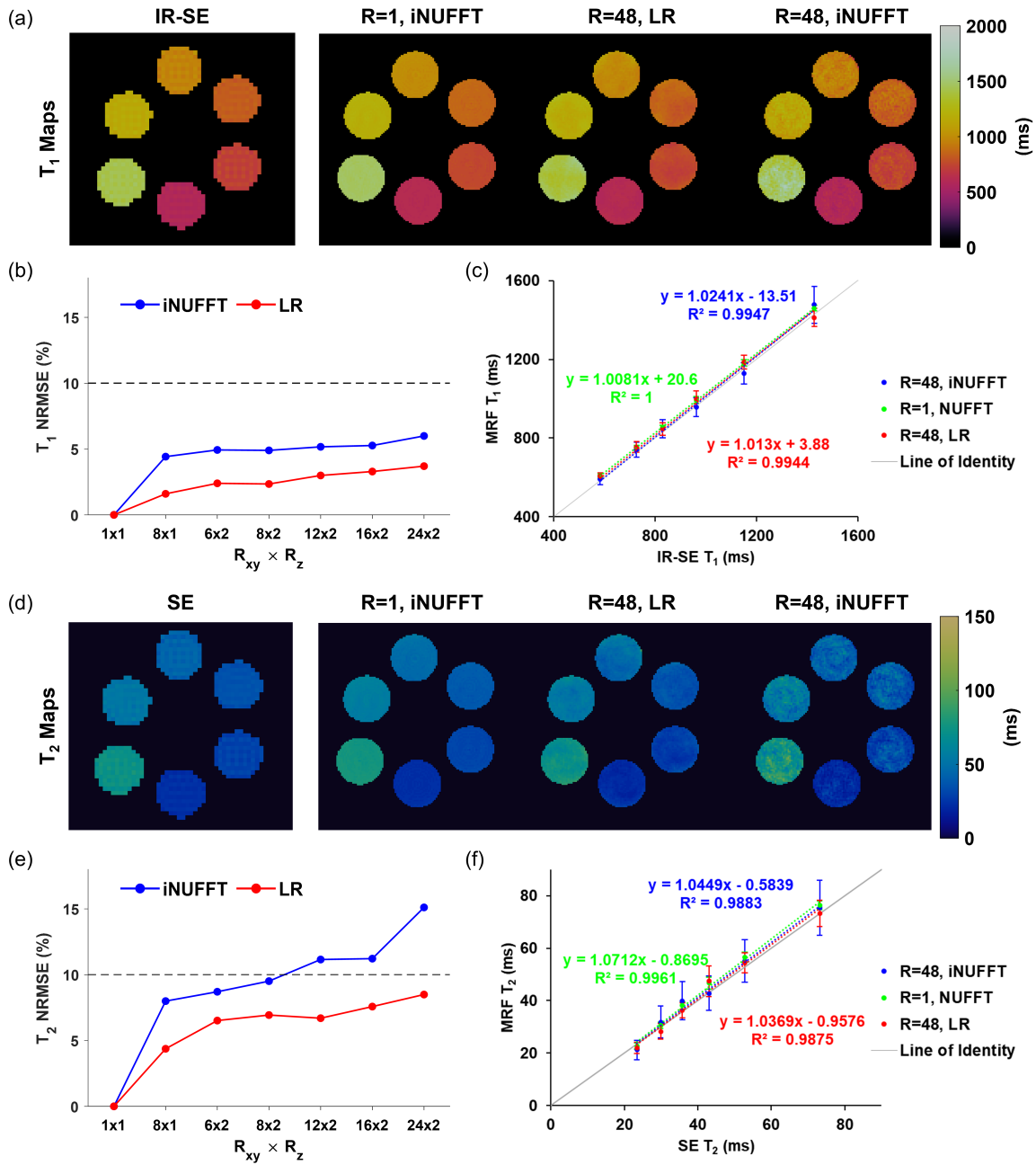

**Figure 2.** In vitro validation. (**a**) T$_1$ maps from inversion-recovery spin-echo (IR-SE) in comparison with fully sampled MRF reconstructed with inverse NUFFT (iNUFFT) and 48-fold undersampled MR fingerprinting (MRF) reconstructed with low-rank (LR) approach and iNUFFT, respectively. (**b**) Normalized root-mean-square error (NRMSE) of T$_1$ maps of a phantom with varied combinations of in-plane (R$_{xy}$) and through-plane (R$_z$) undersampling schemes. (**c**) Correlations between MRF-measured and IR-SE-measured T$_1$ values. (**d**) T$_2$ maps from spin-echo (SE) in comparison with MRF-derived maps. (**e**) NRMSE of T$_2$ maps of a phantom with varied undersampling schemes. (**f**) Correlations between MRF-measured and SE-measured T$_2$ values. The dashed lines in (b) and (e) indicate 10% NRMSE.

Magnetic Resonance in Medicine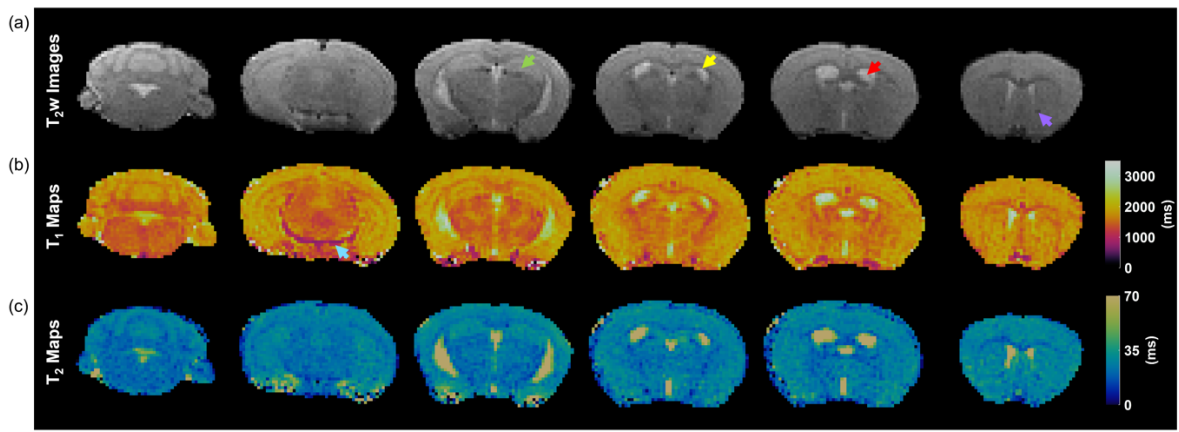

**Figure 3.** $T_1$ and $T_2$ maps of mouse brain by 3D MR fingerprinting (MRF) from axial view. (**a**) $T_2$-weighted images of a representative mouse brain as axial slices from posterior to anterior direction. (**b-c**) Corresponding $T_1$ and $T_2$ maps of the representative mouse brain. Arrows indicate landmark anatomical structures in the brain: Green, hippocampus; Yellow, corpus callosum; Red, ventricular system; Purple, anterior commissure; Blue, posterior cerebral artery.

Page **21** of 31

(a)

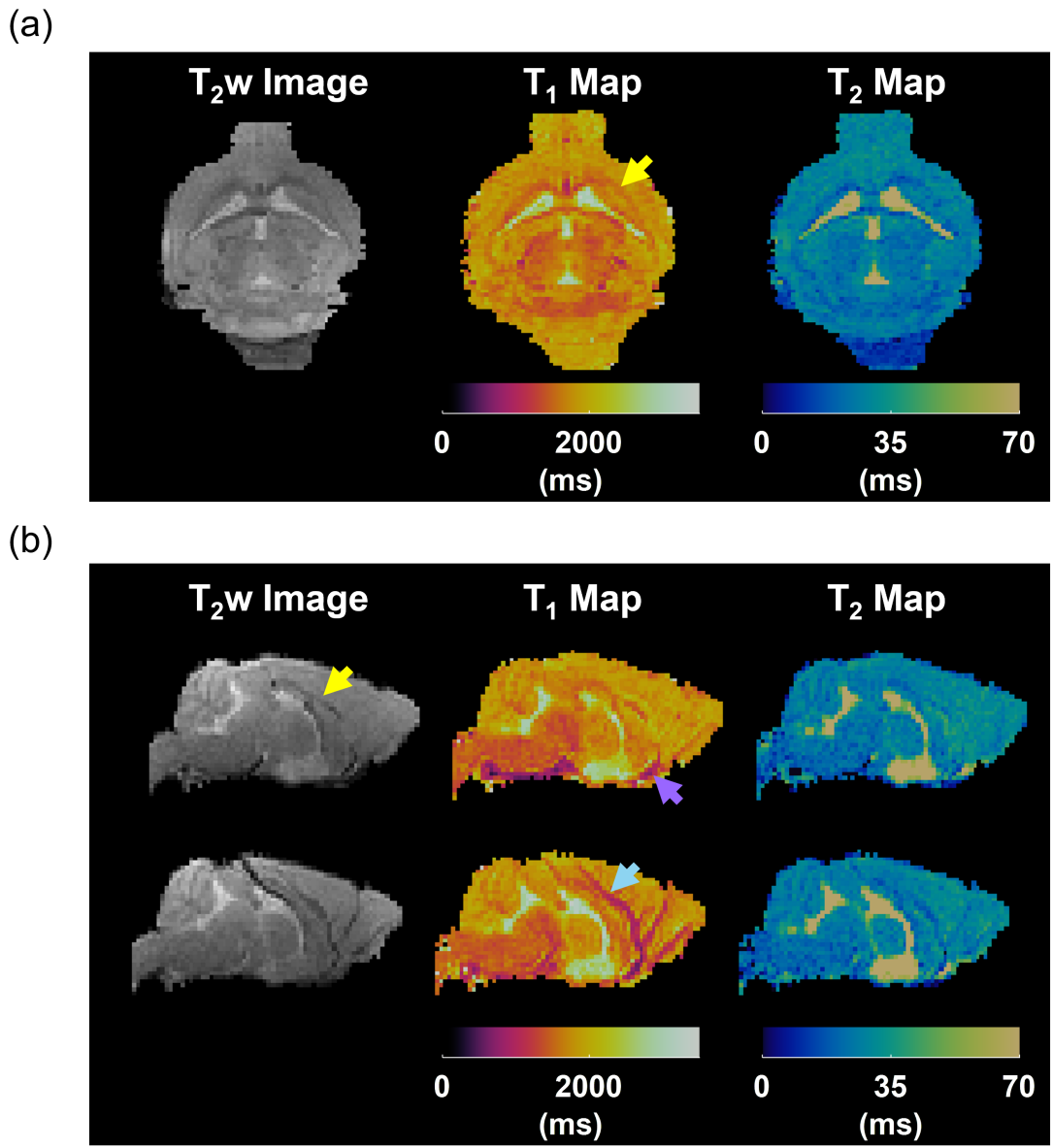

(b)

**Figure 4.** $T_1$ and $T_2$ maps of mouse brain by 3D MR fingerprinting (MRF) from coronal and sagittal views. (**a**) $T_2$-weighted ($T_2$w) image, $T_1$ map, and $T_2$ map of a center coronal slice. (**b**) $T_2$w images, $T_1$ maps, and $T_2$ maps of two center sagittal slices. Arrows indicate landmark anatomical structures in the brain: Yellow, corpus callosum; Purple, anterior cerebral artery; Blue, middle cerebral artery.

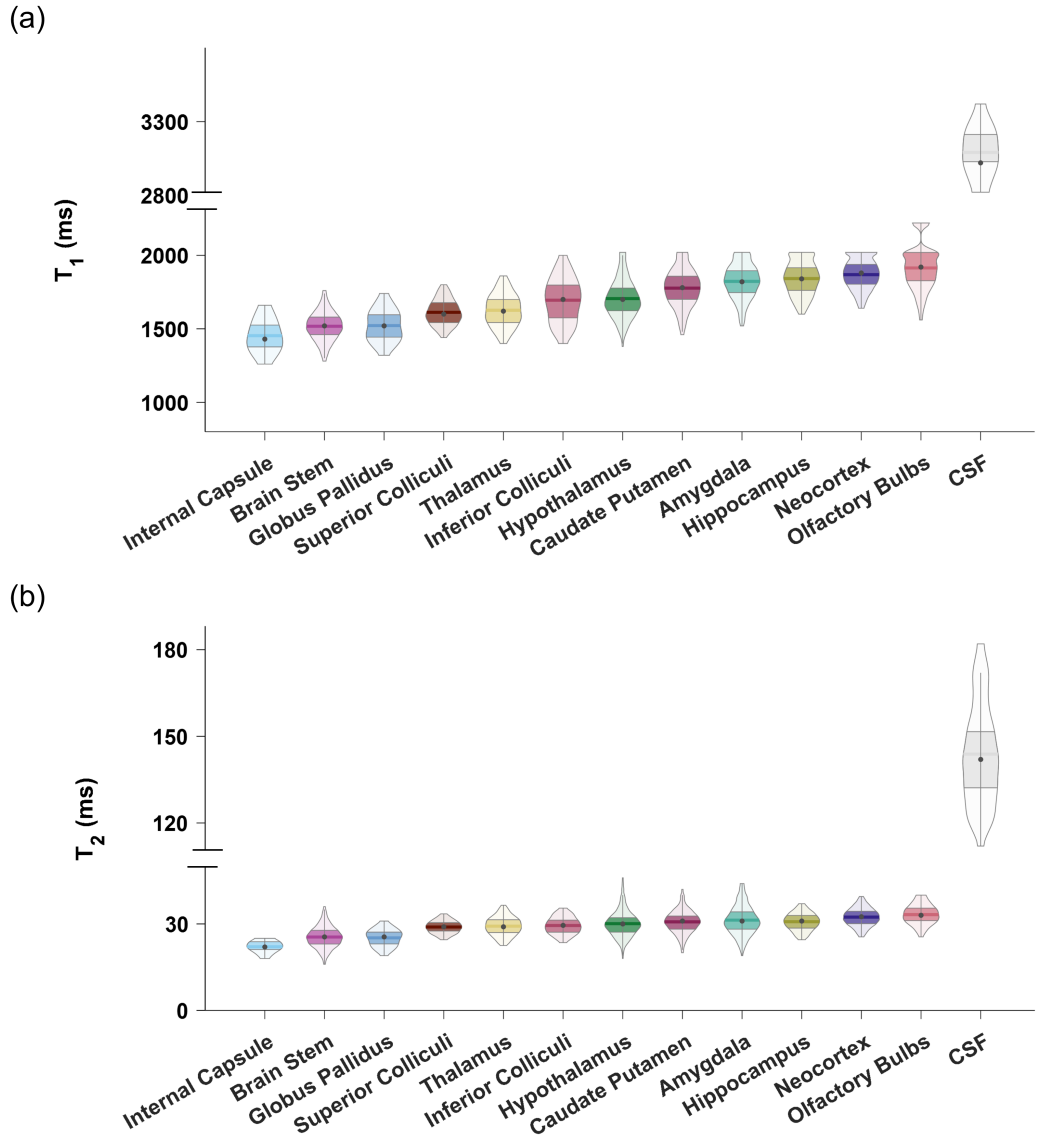

**Figure 5.** Mean $T_1$ (**a**) and $T_2$ (**b**) values in selected regions of interest across the mouse brain (n = 3). Solid dots represent median $T_1$ and $T_2$ values. Solid horizontal lines represent mean $T_1$ and $T_2$ values. CSF, cerebrospinal fluid.

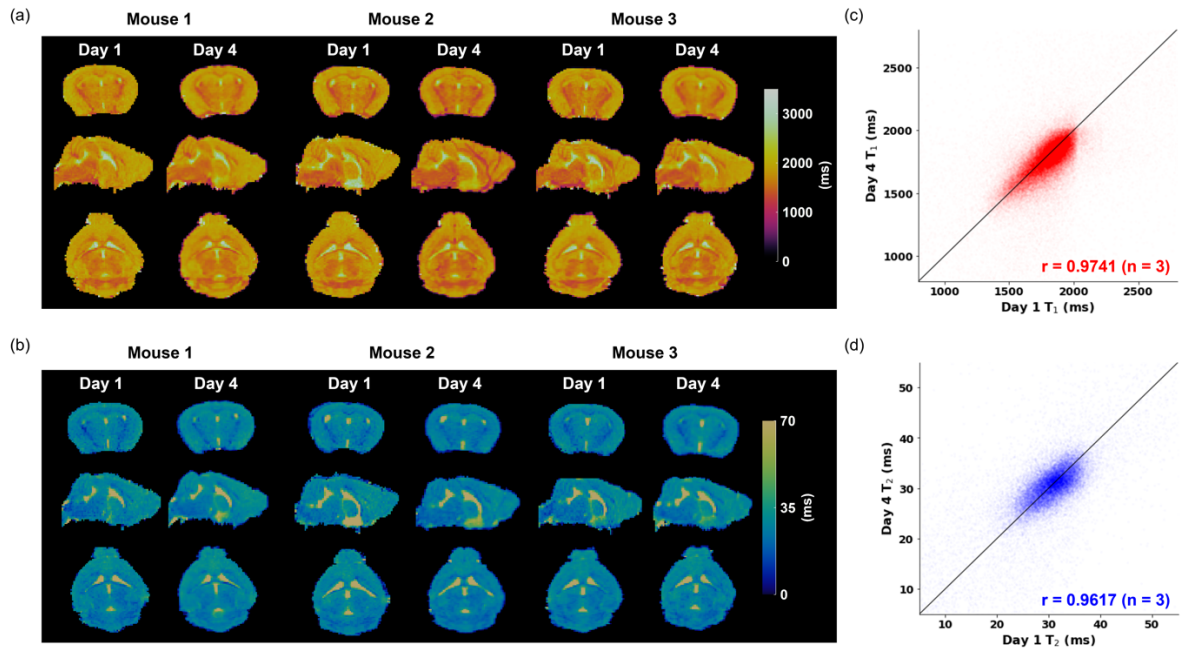

**Figure 6.** Repeatability of in vivo 3D MR fingerprinting (MRF) measurements. (**a-b**) $T_1$ (**a**) and $T_2$ (**b**) maps of three orthogonal slices acquired from two repeated MRF measurements in three animals. The scans were conducted three days apart. (**c-d**) Voxel-wise comparison of $T_1$ (**c**) and $T_2$ (**d**) measurements from two repeated scans of all animals. The black lines represent the lines of identity. r, correlation coefficient.

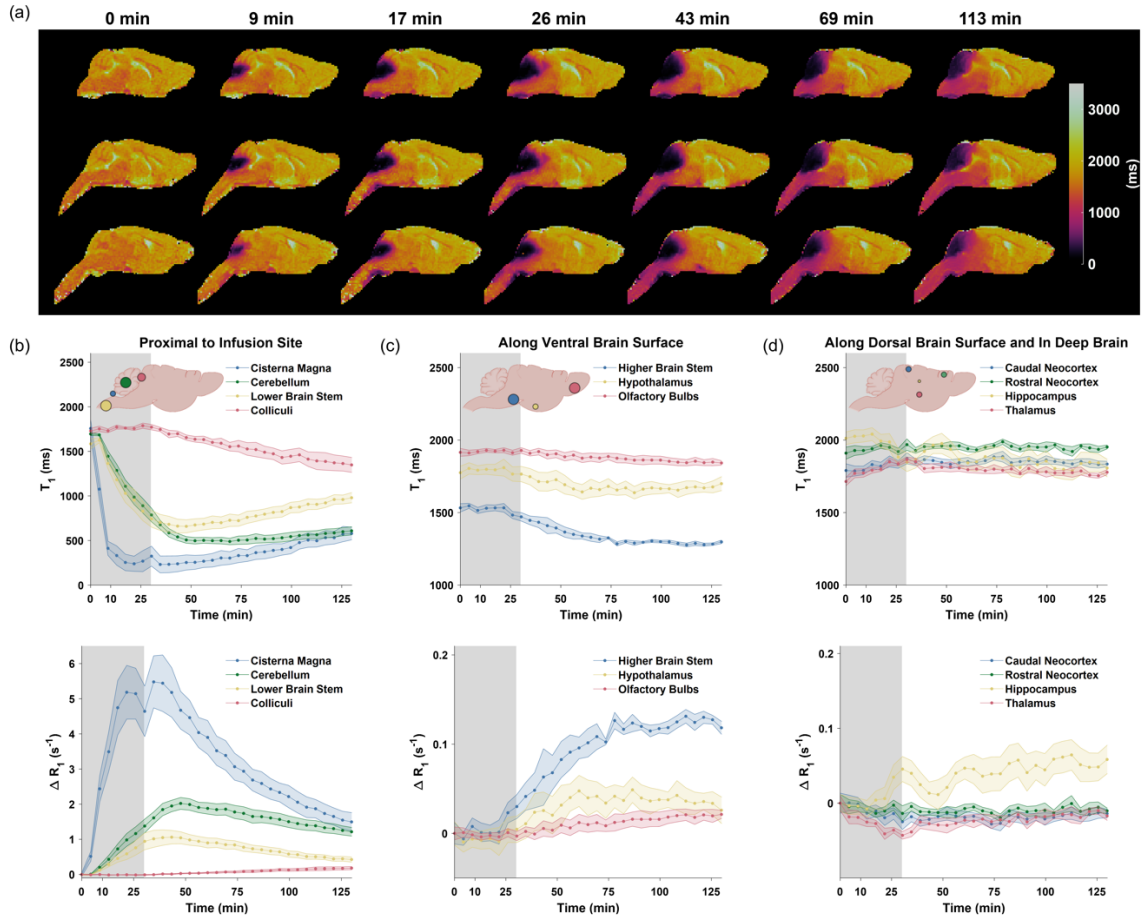

**Figure 7.** Brain-wide dynamic $T_1$ and $\Delta R_1$ changes induced by Gd-DTPA. (**a**) Representative sagittal views of $T_1$ maps overlaid on $T_2$-weighted images from one animal at selected time points. (**b-d**) Time courses of $T_1$ and $\Delta R_1$ changes in selected ROIs. Gray bands indicate the time period of Gd-DTPA infusion. Colored lines represent the mean $T_1$ and $\Delta R_1$ changes (n=5). Shaded areas represent standard errors.

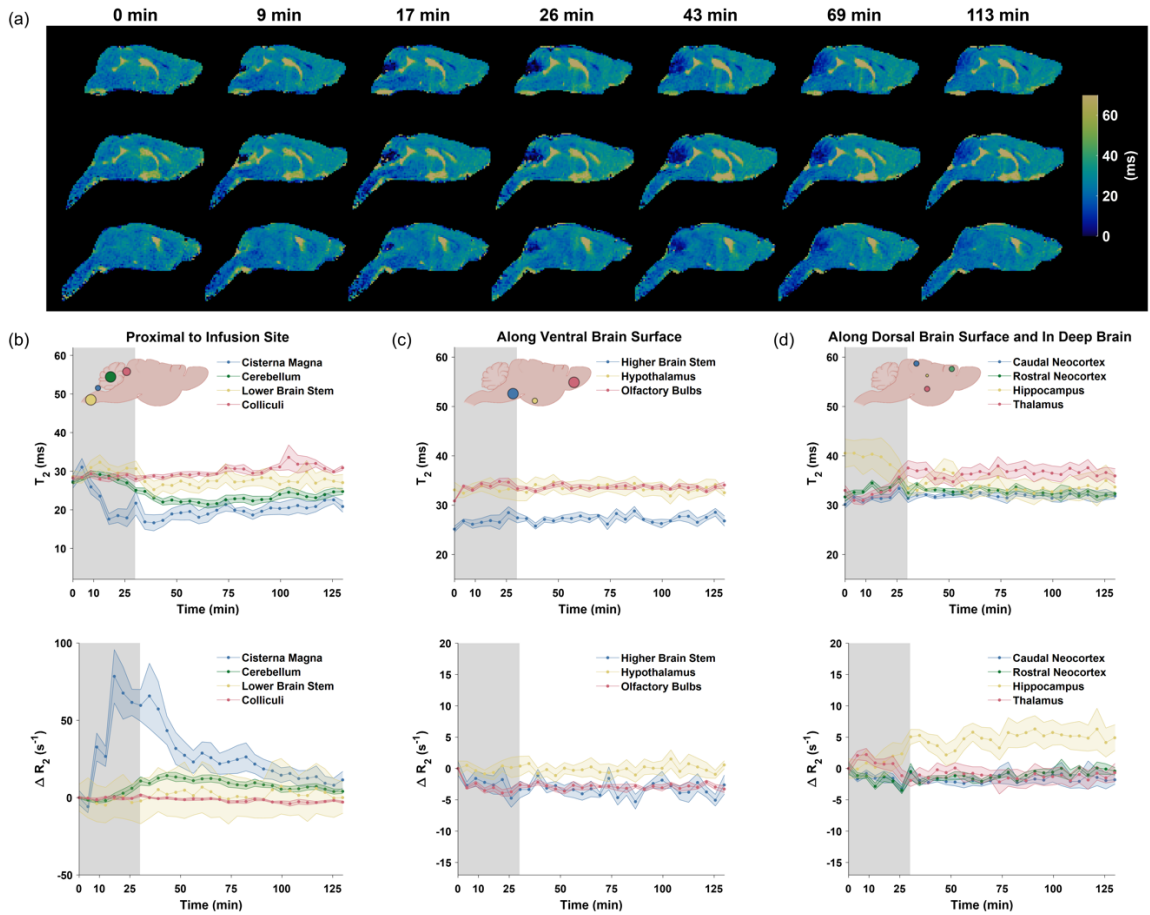

**Figure 8.** Brain-wide dynamic $T_2$ and $\Delta R_2$ changes induced by Gd-DTPA. (**a**) Representative sagittal views of $T_2$ maps overlaid on $T_2$-weighted images from one animal at selected time points. (**b-d**) Time courses of $T_2$ and $\Delta R_2$ changes in selected ROIs. Gray bands indicate the time period of Gd-DTPA infusion. Colored lines represent the mean $T_2$ and $\Delta R_2$ changes (n=5). Shaded areas represent standard errors.

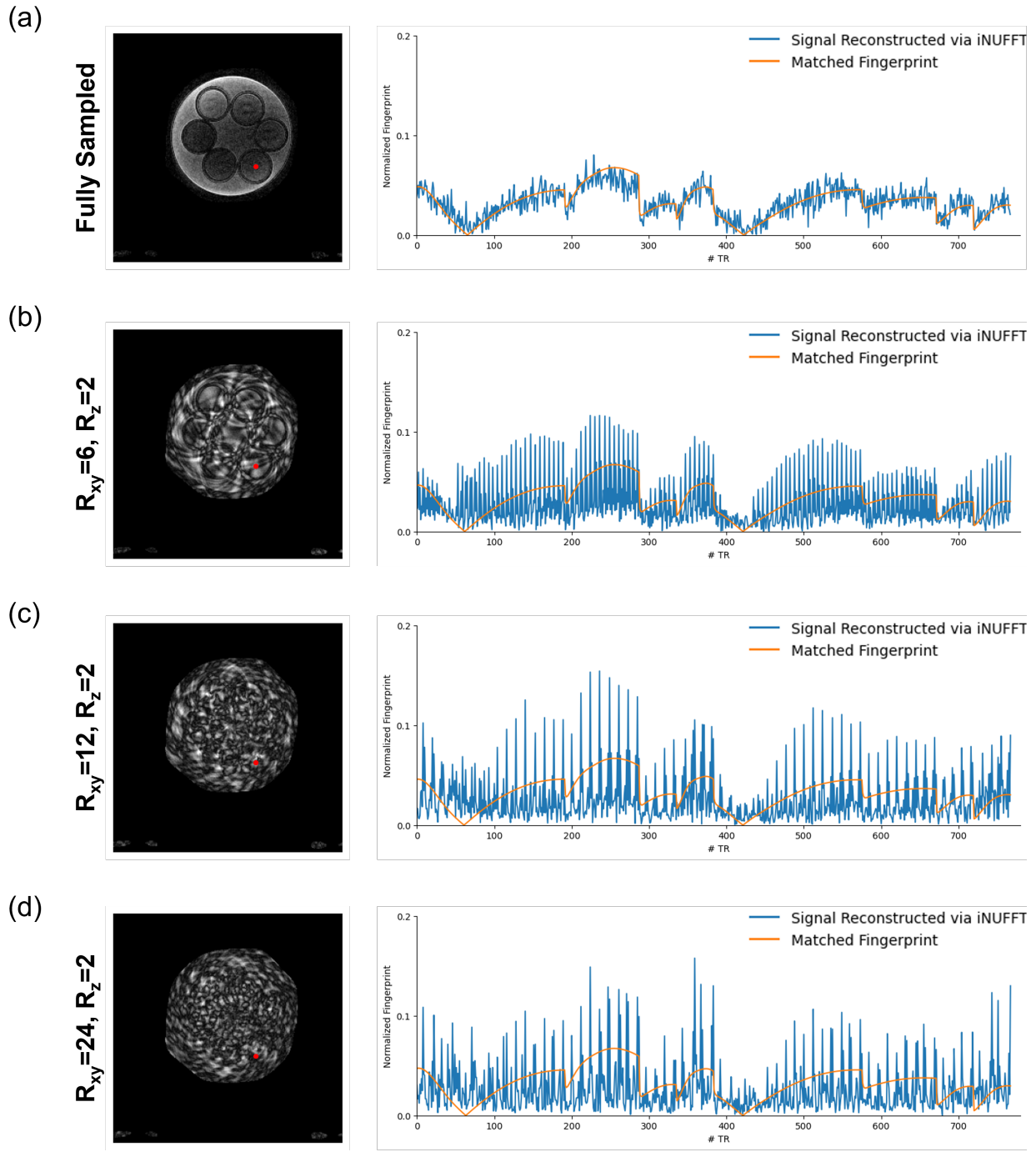

**Figure S1.** Representative images of the 12th frame of an in vitro MRF dataset (left) and the corresponding fingerprints and matched dictionary entries from the voxel highlighted in red (right) from fully sampled (**a**) and undersampled (**b-d**) data. $R_{xy}$ and $R_z$ are the in-plane and through-plane undersampling factors, respectively.

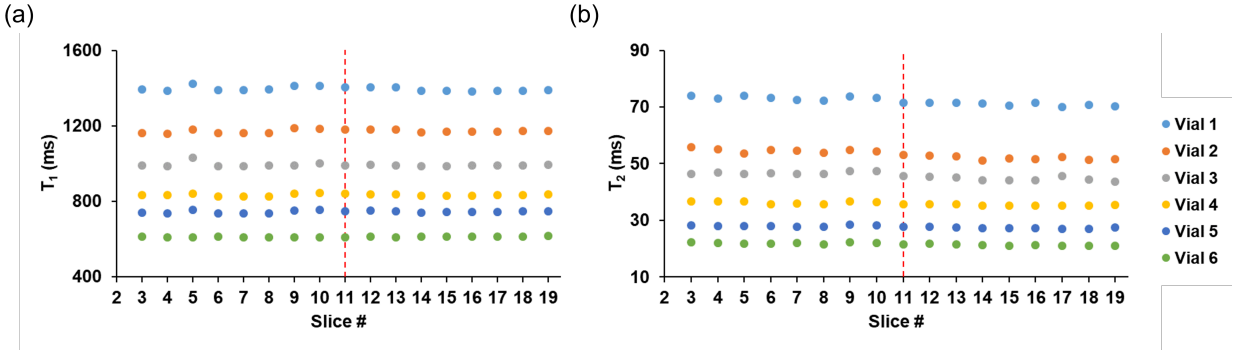

**Figure S2.** Comparison of inter-slice variations in MRF-measured $T_1$ (**a**) and $T_2$ (**b**) values in six phantom vials. Red dotted lines indicate the center slice.

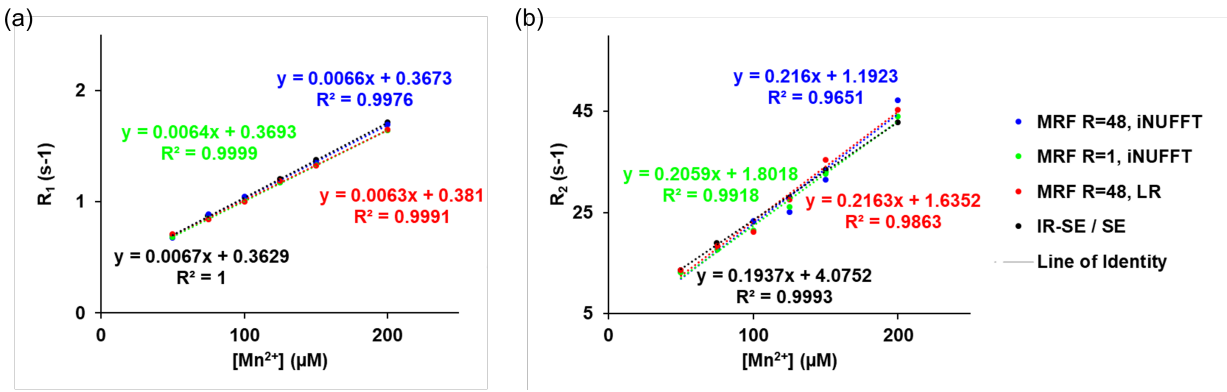

**Figure S3.** Correlation of relaxation rates and $Mn^{2+}$ concentrations. (**a**) Correlation of $Mn^{2+}$ concentration and $R_1$ derived from 3D MR fingerprinting (MRF) and inverse-recovery spin echo (IR-SE) methods. (**b**) Correlation of $Mn^{2+}$ concentration and $R_2$ derived from MRF and spin-echo (SE) methods.

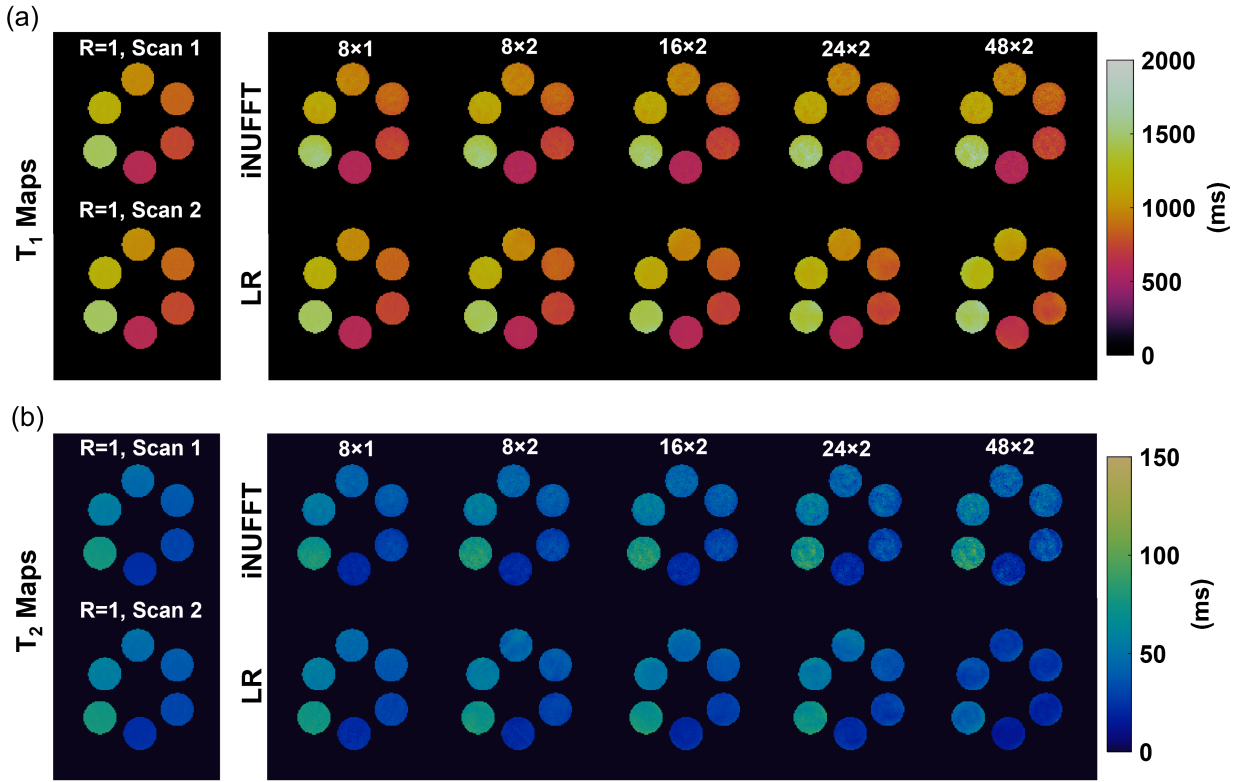

**Figure S4.** $T_1$ and $T_2$ results from fully sampled and undersampled MR fingerprinting (MRF) scans. (**a**) $T_1$ maps from two fully sampled MRF scans and those from undersampled scans with varied in-plane ($R_{xy}$) and through-plane ($R_z$) undersampling factors, reconstructed with inverse non-uniform Fourier transform (iNUFFT) and low-rank (LR) subspace modeling. (**b**) $T_2$ maps from two fully sampled MRF scans and those from undersampled scans with varied $R_{xy}$ and $R_z$ undersampling factors, reconstructed with iNUFFT and LR.

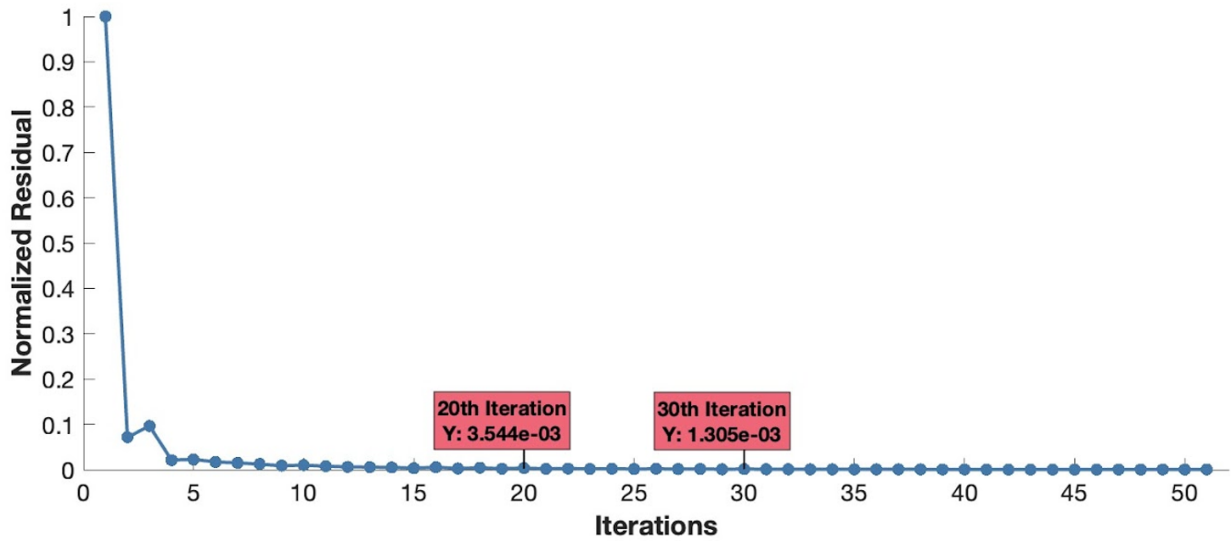

**Figure S5.** Normalized residuals over conjugate gradient (CG) iterations for subspace-based image reconstruction via low rank (LR) for MR fingerprinting (MRF) acquired at 48-fold undersampling. Normalized residual is defined as $\left\|AU\hat{V} - y\right\|/\|y\|$.

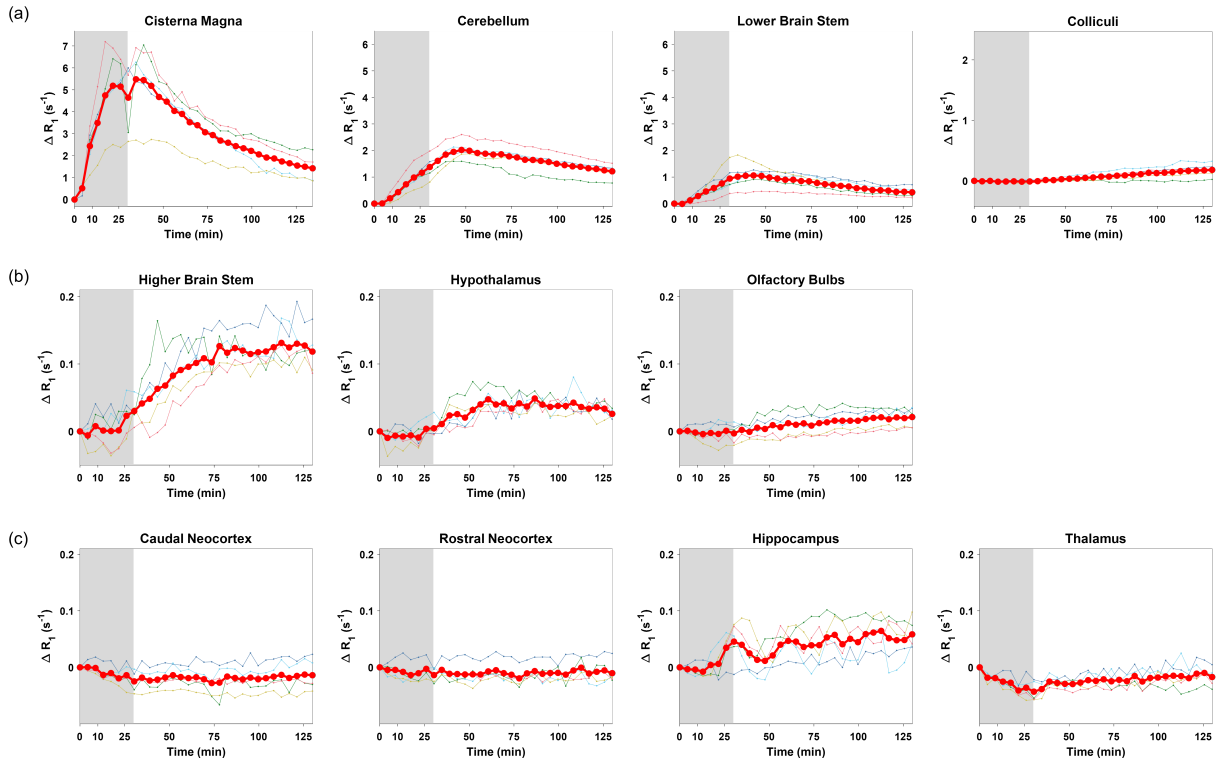

**Figure S6.** Time courses of ΔR$_1$ changes in selected ROIs following Gd-DTPA infusion at cisterna magna. (**a**) Proximal to the infusion site. (**b**) Along the ventral brain surface. (**c**) Along the dorsal brain surface and in the deep brain. The individual animal data (n = 5) are plotted as thin lines, and the mean data are plotted as thick red lines. Shaded areas represent standard errors.

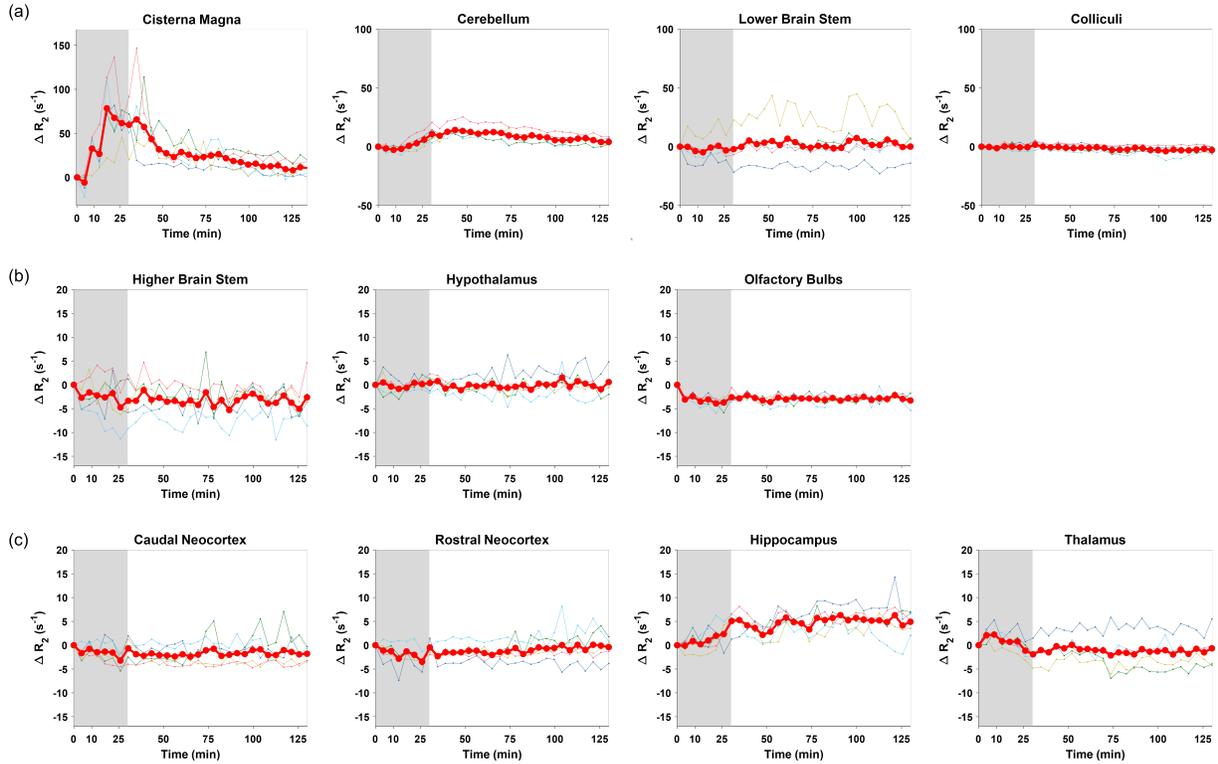

**Figure S7.** Time courses of ΔR$_2$ changes in selected spherical volumes placed at different brain regions following Gd-DTPA infusion at cisterna magna. (**a**) Proximal to the infusion site. (**b**) Along the ventral brain surface. (**c**) Along the dorsal brain surface and in the deep brain. The individual animal data (n = 5) are plotted as thin lines, and the mean data are plotted as thick red lines. Shaded areas represent standard errors.